\documentclass[fleqn,usenatbib]{mnras}

\usepackage{newtxtext,newtxmath}

\usepackage[T1]{fontenc}

\DeclareRobustCommand{\VAN}[3]{#2}
\let\VANthebibliography\thebibliography
\def\thebibliography{\DeclareRobustCommand{\VAN}[3]{##3}\VANthebibliography}

\usepackage{graphicx}	
\usepackage{amsmath}	
\usepackage{multirow}
\usepackage{xcolor}

\newcommand{\source}{4FGL~J0427.8$-$6704}	

\title[Jets in tMSPs]{Jet break revealed in the transitional millisecond pulsar candidate 4FGL~J0427.8$-$6704}

\author[K. I. I. Koljonen et al.]{
K. I. I. Koljonen,$^{1}$\thanks{E-mail: karri.koljonen@ntnu.no}
M. Linares$^{1,2}$,
J. C. A. Miller-Jones$^{3}$
\\
$^{1}$Department of Physics, Norwegian University of Science and Technology, NO-7491 Trondheim, Norway\\
$^2$Departament de F{\'i}sica, EEBE, Universitat Polit{\`e}cnica de Catalunya, Av. Eduard Maristany 16, E-08019 Barcelona, Spain.\\
$^3$International Centre for Radio Astronomy Research, Curtin University, GPO Box U1987, Perth, WA 6845, Australia
}

\date{Accepted XXX. Received YYY; in original form ZZZ}

\pubyear{2015}

\begin{document}
\label{firstpage}
\pagerange{\pageref{firstpage}--\pageref{lastpage}}
\maketitle

\begin{abstract}
Understanding the formation and properties of relativistic jets from accreting compact objects has far-reaching implications in astrophysics. Transitional millisecond pulsars (tMSPs) -- a class of neutron stars transitioning between radio pulsar and accretion states -- offer a unique opportunity to study jet behavior within a low-level accretion regime around fast-spinning, magnetized neutron stars. We analyzed archival spectral energy distributions (SEDs) for both confirmed and candidate tMSPs from literature and various databases, aiming to identify jet spectra and determine physical conditions within these jets. For the tMSP candidate \source, a high-inclination system that displays eclipses in optical, X-ray, and $\gamma$-ray wavelengths, we derived a jet break frequency at $\nu_{\rm br} \approx 10^{11}$ Hz and determined properties of the jet base using a conical jet model (opening angle of $\phi < 32^\circ$, magnetic field strength of $B_{0}\sim100$ G, and radius of $R_{0}\sim10^{10}$ cm). Observations from the Atacama Large Millimeter/submillimeter Array reveal an average flux density of 0.4 mJy, with flares reaching up to 2 mJy on short (seconds) timescales. No eclipses were detected in the millimeter light curves, suggesting the jet base is farther from the central source than in other X-ray binaries ($z_{0}>7\times10^{10}$ cm). We also investigated SEDs of other confirmed and candidate tMSPs but did not find well-defined jet spectral breaks. However, a mid-infrared flux excess in tMSP XSS~J12270$-$4859 suggests that the compact jet emission may extend into near-infrared or optical wavelengths. These results provide new insights into jet formation in tMSPs, highlighting the need for further multi-wavelength observations to fully characterize jet behavior in similar low-accretion systems.
\end{abstract}




\section{Introduction}

The process of accretion onto compact objects and how it leads to the production of relativistic, collimated jets remains a major open question in astronomy. Despite decades of observations, the formation and acceleration of jets from compact objects such as active galactic nuclei, X-ray binaries (XRBs), and gamma-ray bursts is still not fully understood. The energy required to form and accelerate these jets likely originates either from the spin energy of the black hole or neutron star, or from the release of gravitational energy in the accretion disk \citep{blandford82,blandford77}. The kinetic power of these jets remains highly uncertain \citep{fender16} but is crucial for understanding jet formation, the physics of accretion, and the impact of accreting sources on their surroundings through the matter and energy they release. Developing a clear understanding of jet formation from accretion disks around compact objects requires exploring a broad range of parameter space. XRBs span a wide range of accretion rates, providing a unique opportunity to study the evolving properties of jets.

Transitional millisecond pulsars (tMSPs) are an emerging class of neutron stars (with three confirmed pulsars and six candidates) that bridge the gap between XRBs and millisecond radio pulsars in binary systems \citep{papitto22}. These systems alternate between a radio pulsar state and an active, but sub-luminous X-ray disk state, or a brighter accretion/outburst state \citep{papitto13,stappers14,bassa14,linares14}. The transition from the pulsar to the sub-luminous disk state, or from the disk state to the outburst state, occurs rapidly on timescales of days to weeks. However, these sources typically remain in the pulsar or sub-luminous disk state for several years. E.g., PSR~J1023$+$0038 switched to the pulsar state sometime before 2003 \citep{thorstensen05,homer06,archibald09}, and then returned to the sub-luminous disk state ten years later, in 2013 \citep{stappers14,tendulkar14,patruno14}, where it has remained to date. The X-ray luminosities observed in the sub-luminous disk state are relatively low ($L_{\rm X} \sim 10^{33}-10^{34}$ erg/s; 0.5--10 keV) compared to typical XRBs in outburst ($L_{\rm X} \sim 10^{36}-10^{39}$ erg/s), indicating highly sub-Eddington accretion flows or inefficient accretion. During the sub-luminous disk state, tMSPs exhibit two distinct emission modes (high and low, named according to the observed X-ray luminosity; \citealt{linares14,bogdanov15b,archibald15}), which alternate stochastically on timescales of minutes to hours. In addition, some tMSPs (PSR~J1023$+$0038 and candidate sources 4FGL~J0540.0$-$7552 and \source) show rapid flaring behaviour at mm, optical and X-ray wavelengths \citep{tendulkar14,papitto18,strader16,kennedy20,strader21,baglio23}, potentially connected to episodic accretion onto magnetized neutron stars \citep{dangelo10,parfrey16,parfrey17}. However, the average source fluxes remain relatively stable on long (month-year) timescales, and the sub-luminous disk state can persist for years, suggesting a steady but low level of mass accretion towards the MSP \citep{torres17}. 

Interestingly, tMSPs display radio luminosities that are almost as high as those of regular neutron star XRBs in outburst, despite the latter having much higher X-ray luminosities. The enhanced radio luminosity in tMSPs may potentially arise from the interaction between the pulsar magnetosphere and the low-luminosity accretion flow \citep{vandeneijnden22}. Nonetheless, the mechanisms driving outflows from tMSPs remain unclear. Possible explanations include collimated jets \citep{baglio19,baglio23} or outflows launched by the fast-rotating magnetosphere or pulsar wind \citep{papitto14,papitto15,papitto19}. 

A flat radio spectrum has been observed in the tMSP PSR~J1023$+$0038, extending up to at least 100 GHz, similar to what is observed in hard state XRBs. This has been interpreted as evidence for a compact jet \citep{deller15,baglio23}, composed of overlapping, partially self-absorbed synchrotron spectra from various regions within the jet \citep{blandford79}. However, the radio/mm emission from PSR~J1023$+$0038 is highly variable and appears to be anticorrelated with the different X-ray modes: the low mode corresponding to higher radio flux densities, and vice versa \citep{bogdanov18,baglio23}. \citet{bogdanov18} argued that the simultaneous anticorrelation between radio and X-ray modes could result from mechanisms such as rapid plasma discharge from the pulsar magnetosphere or the inflation of a short-lived, compact pulsar wind nebula. The latter interpretation was also supported by \citet{campana19} based on the X-ray properties of PSR~J1023$+$0038 in the different modes. \citet{baglio23} suggested that the compact jet remains continuously active, with the heightened radio flux levels and transient flares in the low mode driven by expulsion of matter from the accretion disk into the jet.

Alternatively, the synchrotron emission could arise from particles accelerated at a termination shock formed by the interaction between the pulsar wind and plasma inflowing from the accretion disk \citep{papitto19,veledina19}. Episodes of magnetic field reconnection at the turbulent termination shock could launch plasmoids that become rarefied and transparent to synchrotron radiation at larger distances, leading to increased radio luminosity \citep{veledina19}.

Interestingly, a similar X-ray/radio anticorrelation pattern has been observed in the tMSP candidate 4FGL J1554.5$-$1126, indicating that the behavior seen in PSR~J1023$+$0038 is not unique \citep{gusinskaia24}. In contrast, the tMSP candidate CXOU J110926.4$-$650224 shows no correlation between X-ray modes and radio emission. In addition, it exhibits higher radio luminosity than PSR~J1023$+$0038 and 4FGL J1554.5$-$1126, suggesting that its radio emission is not connected to the process driving X-ray mode switching, but likely arises from a compact jet \citep{cotizelati21}. These observations highlight the complexity of radio emission mechanisms in tMSPs, suggesting that no single process can fully account for the diverse behaviors observed across sources. This underscores the need for further multiwavelength studies to disentangle these processes and better understand the interplay between accretion, pulsar winds, and jet formation in tMSPs.

In the compact, self-absorbed jet scenario, the flat or inverted jet spectrum is expected to break at some higher frequency, where there is a transition from high to low optical depth (with typical spectral indices above the break $\alpha \sim -0.7$; $F\propto\nu^{\alpha}$). This transition is thought to occur either at the jet base \citep[e.g.,][]{blandford79,konigl81,ghisellini85}, or at the location where particles are accelerated in a shocked zone \citep[e.g.,][]{markoff05,marscher08,polko14}. However, due to the limited availability of multiwavelength observations, especially in the sub-mm/mm and far/mid-infrared bands, the jet spectral break has only been inferred in a few neutron star XRBs, with a wide range of frequencies from $10^{10}$ Hz to 10$^{15}$ Hz \citep{migliari10,baglio16,diaztrigo17,diaztrigo18}. It is still unclear whether the jet spectral breaks in neutron XRBs differ systematically from those in black hole XRBs, which typically exhibit a mean break frequency of $\nu_{\rm br} \sim 10^{14}$ Hz during the hard state, when compact jets are present \citep[e.g.,][]{russell13,koljonen15}.

In this paper, we examine archival spectral energy distributions (SEDs) of confirmed and candidate tMSPs from the literature and various databases, with the goal of identifying jet spectra and to derive the physical conditions within the jet. In Section \ref{sec:obs}, we describe the data used, focusing on the tMSP candidate \source\/ with the best available spectral coverage. In Section \ref{sec:res}, we present detailed analysis of Atacama Large Millimeter/submillimeter Array (ALMA) data for \source\/ and SED modeling. In Section \ref{sec:discuss}, we discuss the implications of these results for the jet properties of \source, and compare them to other systems. We conclude in Section \ref{sec:conclude}. In Appendix A, we review the SEDs of two confirmed tMSPs (XSS J12270$-$4859 and PSR J1023+0038) and another candidate (4FGL J1544.5$-$1126) with short descriptions of their data sets and fitting results.

\section{Observations} \label{sec:obs}

We compiled the SEDs of confirmed tMSPs and candidate sources from the literature. The most comprehensive datasets are available for the tMSPs PSR~J1023$+$0038, XSS~J12270$-$4859, and the candidates 4FGL~J1544.5$-$1126 and \source. We obtained the radio data of these sources from several papers, as well as unpublished ALMA observations of \source, which we analyze here. In addition, we retrieved infrared, optical and ultraviolet photometric data for these four tMSPs or candidates from the VizieR Catalogue. X-ray data were obtained through the High Energy Astrophysics Science Archive Research Center (HEASARC). In the following section, we describe the data for \source; however, the compiled SEDs for PSR~J1023$+$0038, XSS~J12270$-$4859, and 4FGL~J1544.5$-$1126 are detailed and shown in Appendix A.

\subsection{4FGL~J0427.8$-$6704}

\source\/ is a candidate tMSP, observed so far only in a sub-luminous disk state. It exhibits eclipses in its optical, X-ray, and $\gamma$-ray light curves, indicating a high-inclination orbit for the binary system \citep[$84 \pm 3^{\circ}$;][]{kennedy20}. The distance to \source, estimated from its \textit{Gaia} DR3 parallax \citep{gaia16,gaia22}, is 2.5$^{+0.5}_{-0.3}$ kpc \citep{koljonen23}, consistent with the distance estimate (2.4 kpc) from optical light curve modeling by \citet{strader16}. The flat spectrum, stable mean radio luminosity of $L_{\rm 5 GHz} \sim 10^{28}$ erg s$^{-1}$ -- comparable to the jet luminosities of black hole XRBs at similar X-ray luminosities -- and the absence of eclipses, which suggest emission originating farther from the neutron star, collectively point to the radio emission arising from a compact jet. However, other scenarios mentioned in Section 1 cannot be definitively ruled out.

Although the soft X-ray light curve of \source\/ does not show the clear mode switching characteristic of tMSPs, it displays flaring behavior, with an average X-ray luminosity of $(2.6\pm0.2) \times 10^{33}$ erg s$^{-1}$ in the soft X-ray band \citep[0.5--10 keV;][]{strader16} and varying between (0.5--20) $\times 10^{33}$ erg s$^{-1}$, with the highest luminosities measured during flares occurring on average every half-hour \citep{li20}. This luminosity range closely resembles that of PSR J1023$+$0038 in the sub-luminous disk state, which shows average luminosities of 0.5 $\times 10^{33}$ erg s$^{-1}$ in the low mode, 3 $\times 10^{33}$ erg s$^{-1}$ in the high mode, and up to 11 $\times 10^{33}$ erg s$^{-1}$ during flares \citep{bogdanov15b}. The soft X-ray spectra of \source\/ are best-fit with a partially absorbed power law model, with photon indices ranging from $\Gamma=1.2-1.8$ and column densities of $N_{\rm H}\sim10^{23}$ atoms cm$^{-2}$, along with high covering fractions over 0.96 \citep{li20,kennedy20}. The partial absorption likely originates within the system and is attributed to our edge-on view of the accretion disk, which obscures regions near the neutron star \citep{kennedy20}. The obtained photon indices are also consistent with other tMSP systems, which generally exhibit $\Gamma=1.6-1.8$ \citep{demartino13,tendulkar14,linares14,bogdanov15b}.

Additional evidence supporting the tMSP classification includes the system's short orbital period of 8.8 hours, component masses derived from light curve modeling ($M_{\rm NS} \approx 1.4 M_{\odot}$, $M_{\rm comp} \approx 0.3 M_{\odot}$), $\gamma$-ray emission originating very close to the compact object (as suggested by persistent eclipses), and relatively strong radio emission of 0.3 mJy \citep{strader16,kennedy20,li20}. 

We use the radio data described in \citet{li20}, which were obtained with the Australia Telescope Compact Array (ATCA) on May 2nd and 3rd, 2017. The radio data do not show any indication of eclipses, and the mean flux density was measured as 0.290$\pm$0.007 mJy at 5 GHz and 0.300$\pm$0.006 mJy at 9 GHz, with a mean spectral index of $\alpha = 0.07\pm0.07$ indicating a flat radio spectrum.

In addition, we reduced and analyzed three ALMA observations (detailed below) taken in December 2017 and May 2018, as well as one NuSTAR observation from May 2016. We futher supplemented the SED with infrared, optical, and ultraviolet data from the Vizier Catalogue and \citet{strader16}. 

\subsubsection{ALMA}

\begin{figure*}
  \centering
  \includegraphics[width=0.49\textwidth]{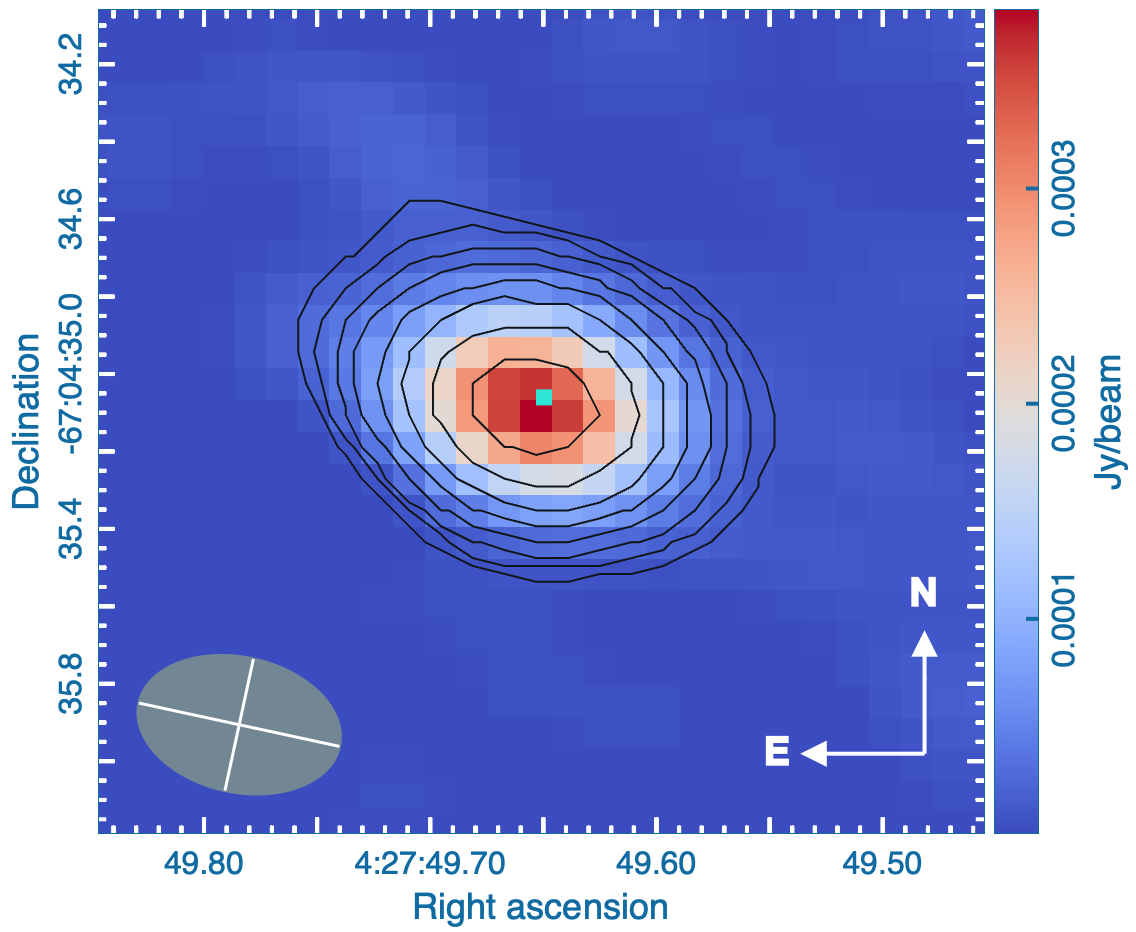}
  \includegraphics[width=0.49\textwidth]{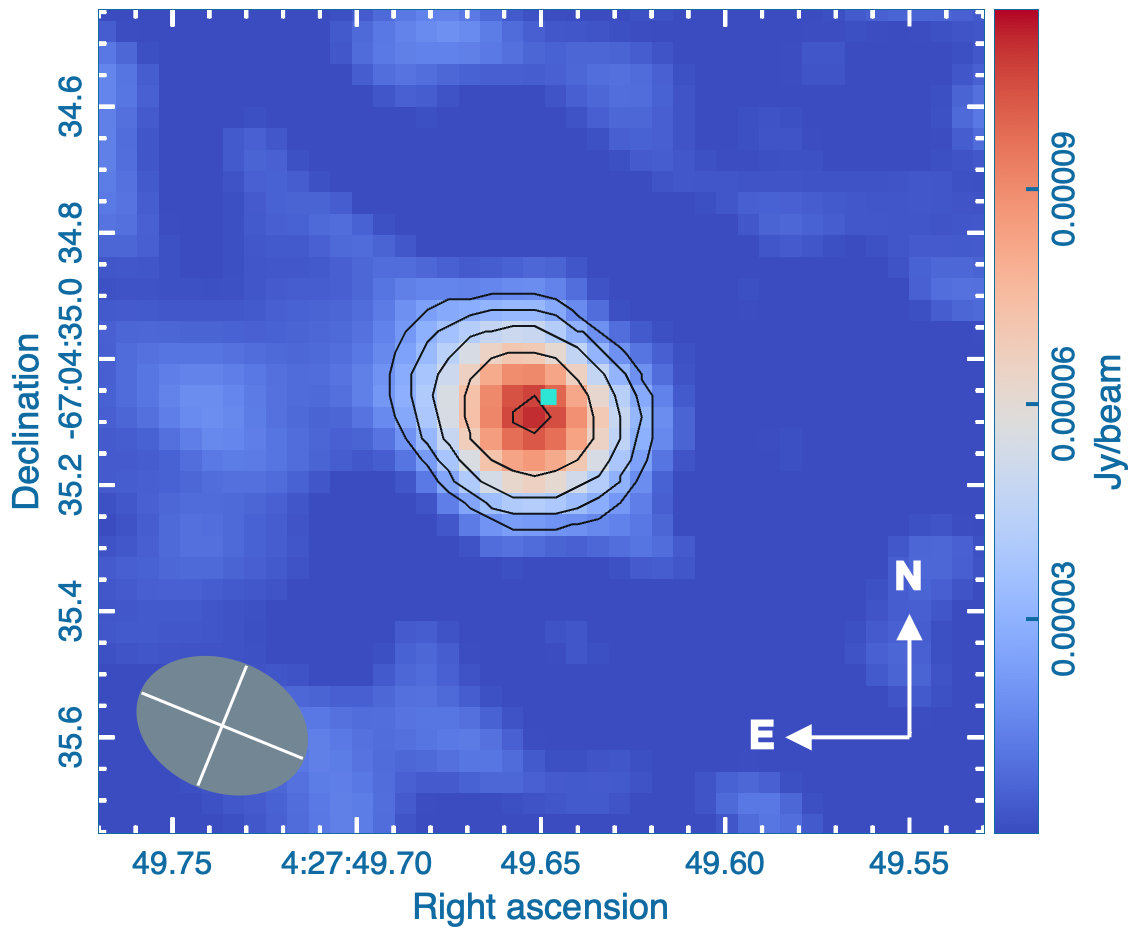}
  \caption{The ALMA images of \source\/ in Band 3 (left) and Band 6 (right). The contours show flux levels with $\sqrt{2}^n\times\sigma_{\mathrm{rms}}$, where $n=1,2,3\ldots8$. The synthesized beams are shown in the bottom left. The \textit{Gaia} location is marked as a filled cyan square.}
  \label{fig:alma}
\end{figure*}

For \source, there are three ALMA observations available in the ALMA archive, obtained on 21-Dec-2017 (Band 3; 97.5 GHz), 28-Dec-2017 (Band 6; 233 GHz), and 17-May-2018 (Band 6). All observations were conducted with the 12-m full array using 48, 45, and 43 antennas, respectively, with configuration C43-6 for the first two epochs and C43-2 for the third. The corresponding baseline ranges were 15-2500 m fo the first two epochs and 15-314 m for the third.   

We acquired the calibrated measurement sets from the European ALMA Regional Center. We flagged data as suggested in the qualification reports and performed a manual CLEAN using CASA 6.5.4.9. We applied a Briggs robust parameter of 2 (i.e., natural weighting) for optimal sensitivity, and 0 for improved angular resolution, using a Hogbom deconvolver. We produced images for the total band, per science window, and on a intra-observation basis (per scan or shorter time step, depending on the source brightness). 

The source was detected at the 33$\sigma$ level in Band 3 and at the 11$\sigma$ level in Band 6 (9$\sigma$ in the May 2018 observation). The average flux densities were measured as 0.388$\pm$0.009 mJy in Band 3 and 0.114$\pm$0.008 mJy, and 0.152$\pm$0.013 mJy in the two Band 6 observations. However, as we show below, the intra-observation flux density is highly variable. Fig. 1 presents the total band ALMA images from the first two observing epochs. These images can be fitted with a 2D Gaussian models whose parameters match those of the beam, indicating that the source is unresolved. No additional sources or extended structures are visible within the field of view (60 arcsec for Band 3). The Band 6 images taken with configuration C43-6 and using robust=0 weighting offer the highest resolution, setting an upper limit on the source size to 0.1 arcsec (major axis) and 0.05 arcsec (minor axis). 

We estimated the peak flux densities from the 2D Gaussian fits using CARTA 4.1.0, fixing the Gaussian minor and major axes, as well as the position angle, to match the beam parameters. We also used the UVMULTIFIT routine in CASA to estimate the flux densities directly from the u-v plane, assuming a point source. These estimates are consistent with the peak flux densities obtained from the imaging. 

The mm source is localized at RA 04:27:49.6530 $\pm$ 0.0004, Dec $-$67:04:35.053 $\pm$ 0.003 in Band 3, and RA 04:27:49.6530 $\pm$ 0.0006, Dec $-$67:04:35.060 $\pm$ 0.006 in Band 6 images (statistical errors only). The \textit{Gaia} DR3 counterpart is located at RA 04:27:49.64627, Dec $-$67:04:35.0611 (epoch 2016.0). Given the source's high proper motion (12.5 mas/yr in RA, 0.4 mas/yr in DEC), the corrected \textit{Gaia} position at the time of the first two ALMA observations (epoch 2018.0) is RA 04:27:49.64794 DEC -67:04:35.0603 (marked as a cyan square in Fig. \ref{fig:alma}). The corrected \textit{Gaia} position is 30 mas from the Band 3 and Band 6 source centers. However, the nominal positional accuracy of ALMA in these observations is $\sim$30 mas\footnote{This depends on the FWHM of the synthesized beam, the source's signal-to-noise ratio, and a factor for signal decorrelation, as described in the ALMA technical handbook.}. Thus, the ALMA and \textit{Gaia} locations agree after correcting for proper motion. 

\subsubsection{NuSTAR}

The only \textit{NuSTAR} observation available for \source, taken on 19-May-2016, is analyzed in \citet{strader16}, to which we refer the reader for full details. Briefly, the average countrate for the 60 ks on-source exposure time was (0.049 $\pm$ 0.001) counts s$^{-1}$ per detector, with strong variability between zero counts (eclipses) and up to 0.1 counts s$^{-1}$. The average spectrum over the detector's energy band (3--79 keV) is well fitted by an absorbed power law model, with a column density of $N_{\rm H} = (6\pm2) \times 10^{22}$ atoms cm$^{-2}$, a photon index of $\Gamma=1.7\pm0.1$, and an X-ray flux of (6.8$\pm$0.4) $\times 10^{-12}$ erg s$^{-1}$ cm$^{-2}$, corresponding to a luminosity of (5$\pm$0.3) $\times 10^{33}$ erg s$^{-1}$ at 2.5 kpc.

To facilitate SED modeling, we reduced the NuSTAR observation of \source\/ and extracted the average spectrum. The data from both focal plane modules (FPMA and FPMB) were processed using \textsc{nupipeline} in \textsc{HEAsoft}. We used a circular source region with an 80-arcsec radius centered on the source location, and circular background region with a 160-arcsec radius selected from a source-free area in the detector image. We extracted an averaged spectrum from both detectors using entire observation period. For SED modeling, we used the FPMA spectrum, heavily binning the data to a minimum signal-to-noise ratio (S/N) of 15 in the 3--10 keV band and to S/N ratio of 10 above 10 keV band to reduce the number of data points and their associated error, improving fitting stability.

\section{Results} \label{sec:res}

\subsection{mm variability}

\begin{figure*}
 \centering
 \includegraphics[width=\linewidth]{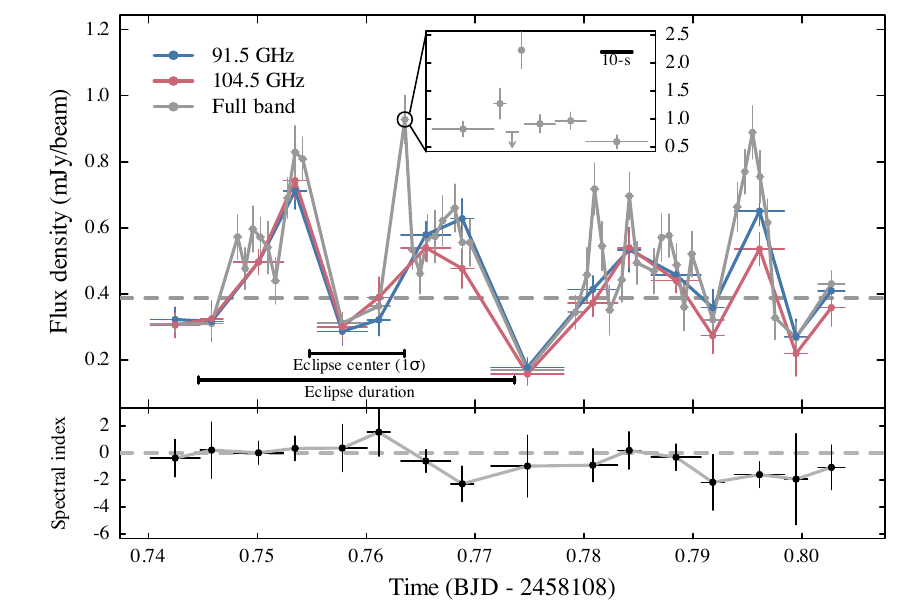}
 \caption{\textit{Upper panel:} Intra-observation light curve of \source\ from the ALMA Band 3 observation. The blue and red lines represent the sidebands with central frequencies of 91.5 GHz and 104.5 GHz, respectively, with time bins ranging from 3 to 10 minutes. The grey line indicates the full band, with time bins ranging from 1 to 10 minutes. The dashed grey line denotes the average flux density of the whole observation. The inset shows the brightest 1-min time bin analyzed at a finer time resolution (down to 2-sec). The predicted center of the eclipse, along with its 1$\sigma$ uncertainty range, and the eclipse duration are shown by the solid line segments. \textit{Lower panel:} The spectral index evolution in the sidebands.}
 \label{fig:ob1_lc}
\end{figure*} 

\begin{figure}
 \centering
 \includegraphics[width=\linewidth]{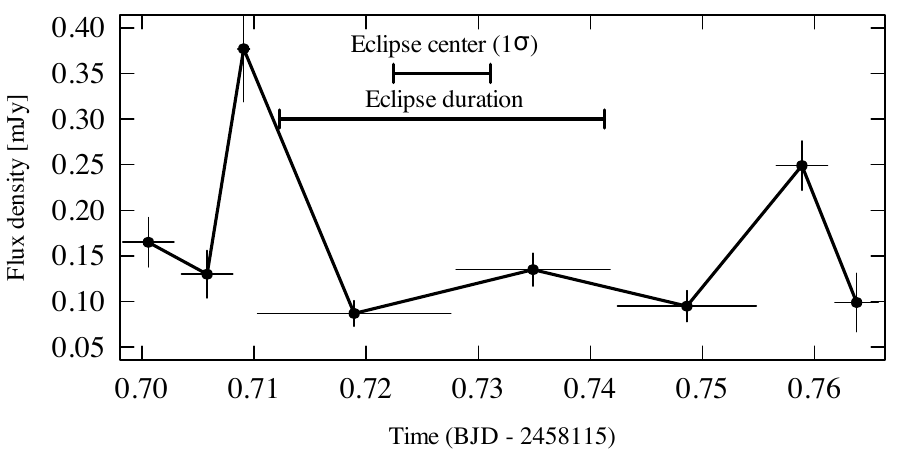}
 \includegraphics[width=\linewidth]{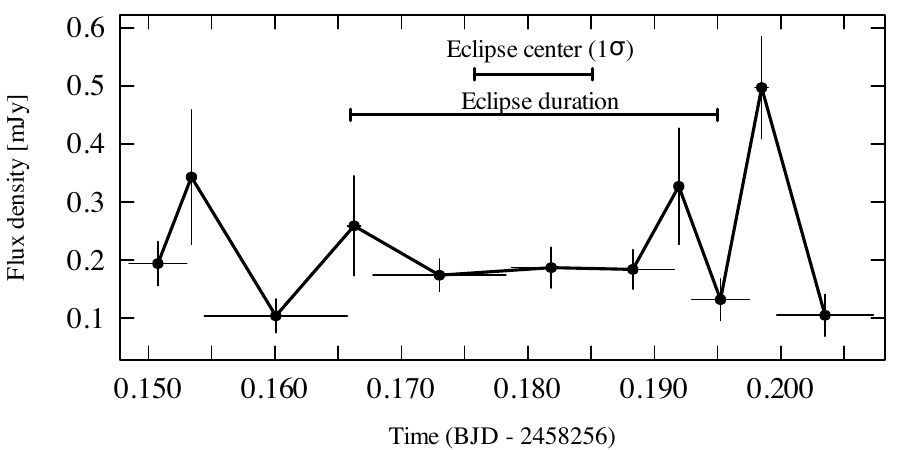}
 \caption{Intra-observation light curve of \source\ from the ALMA Band 6 observations. Due to lower brightness, only total band flux densities are shown, with time bins ranging from 1.5 to 20 minutes. The predicted centers of the eclipse, along with their 1$\sigma$ uncertainty range, and the eclipse duration are shown by the solid line segments.}
 \label{fig:ob23_lc}
\end{figure} 

Fig. \ref{fig:ob1_lc} shows the intra-observation light curves from the Band 3 observation. The time bins for the sidebands (blue and red lines in Fig. \ref{fig:ob1_lc}, corresponding to sideband central frequencies of 91.5 GHz and 104.5 GHz) range from 3 to 10 minutes, while the time bins for the total band (grey lines) range from 1 to 10 minutes. These were adjusted to ensure that the source significance remained above 4$\sigma$ at all times, which allow us to probe the highest time resolution possible at a given time. In addition, we examined the brightest 1-min time bin at a finer resolution, down to 2-sec intervals (see inset in Fig. \ref{fig:ob1_lc}). Fig. \ref{fig:ob23_lc} presents the intra-observation light curves from the Band 6 observations. However, due to the lower brightness, only the total band flux densities are shown, with time bins ranging from 1.5 to 20 minutes. We have converted the time bins into a solar system barycenter time (BJD).   

The mm flux density is clearly variable with flaring occurring on minute timescales. The flares reach flux densities of 0.9 mJy/beam in Band 3 and 0.5 mJy/beam in Band 6. This variability is reminiscent of the X-ray and optical flaring behavior, where the light curves are dominated by flares with timescales of 10--100 seconds \citep{li20}. The fastest flare observed in Band 3 lasts two seconds, reaching a flux density of $\sim$2.0 mJy/beam, brightening by at least a factor of three. The light crossing time for the emission region of this flare is $R_{0} < ct$(2 sec) = 6$\times 10^{10}$ cm, setting an upper limit on the photosphere size at 97.5 GHz. In addition, we can estimate a lower limit on the brightness temperature of this region as $T_{b}=S_{\nu}c^2/2k\Omega\nu^2>7\times10^9$ K, where $S_{\nu}$ is the flux density (taken as the average), $k$ is the Boltzmann's constant, $\Omega$ is the solid angle (estimated from the photosphere size at the source distance), and $\nu$ is the emission frequency. The high brightness temperature implies that the observed mm emission is inherently non-thermal in nature.

Based on Fig. \ref{fig:ob1_lc}, the Band 3 spectrum appears to be variable, with spectral flattening or turning to inverted occurring near the flare peaks. This behavior is similar to that of a typical XRB jet, where optically thick blobs or shocks become optically thin as they travel within the jet. 

We do not observe any clear eclipses in the mm light curves. In Figs. \ref{fig:ob1_lc} and \ref{fig:ob23_lc}, we mark the predicted eclipse center times as derived from \citet{kennedy20}: $T_{n} = T_{0} + P*n = (2455912.83987\pm0.00095) + (0.3667200\pm0.0000007) \times n$, with $n=5988, 6007, 6390$ for each ALMA observing epoch. The associated errors on $T_{n}$ are propagated accordingly, and the1$\sigma$ uncertainty ranges are shown as solid line segments in both figures). 

The eclipse duration in the $\gamma$-ray, X-ray, and optical regimes is 42 minutes ($\sim$0.03 days; also indicated in the figures). However, the Band 3 mm light curve does not display any such low-flux density phase. Notably, the center of the lowest flux density bin (at BJD 2458108.775) deviates by more than 3$\sigma$ from the predicted eclipse center. In the ALMA Band 3 light curve, the eclipse center coincides with a local minimum (Fig. \ref{fig:ob1_lc}). However, the flux density at this point is less than 2$\sigma$ from the average. Moreover, there are four other minima in the light curve with equally low or lower flux densities, making this coincidence with the eclipse center rather unsurprising. Similarly, no corresponding low-flux density phase is observed in the Band 6 light curves (Fig. \ref{fig:ob23_lc}). Therefore, we conclude that the bulk of the mm emission is not eclipsed.
 
\subsection{mm spectrum}

The averaged spectrum in the Band 3 observation is flat or slightly optically thin, with a slope of $-$0.35$\pm$0.38. The average flux densities in the individual spectral windows are 0.427$\pm$0.016 mJy/beam at 90.5 GHz, 0.396$\pm$0.016 mJy/beam at 92.4 GHz, 0.362$\pm$0.015 mJy/beam at 102.5 GHz, and 0.414$\pm$0.016 mJy/beam at 104.5 GHz.  

The averaged spectra in the Band 6 observations are similarly flat or slightly inverted, with a slope of $-$0.22$\pm$0.55 and $-$0.04$\pm$0.58 for the second and third epochs, respectively. The peak flux densities in the individual spectral windows are 0.126$\pm$0.014 mJy/beam at 224 GHz, 0.121$\pm$0.014 mJy/beam at 226 GHz, 0.096$\pm$0.014 mJy/beam at 240 GHz, and 0.117$\pm$0.018 mJy at 242 GHz for the observation on 28-Dec-2017. For the 17-May-2018 observation, the peak flux densities are 0.155$\pm$0.025 mJy/beam at 224 GHz, 0.164$\pm$0.023 mJy/beam at 226 GHz, 0.155$\pm$0.026 mJy/beam at 240 GHz, and 0.136$\pm$0.028 mJy at 242 GHz. 

Thus, the in-band spectrum is flat or slightly inverted. However, the spectral slope from Band 3 to Band 6 is notably steeper, with the averaged flux densities of 0.39 mJy in Band 3 and 0.15/0.11 mJy in Band 6 corresponding to $\alpha$=[-1.1/-1.4]$\pm$0.1. While these observations are not simultaneous, the consistency between the average flux densities in the two Band 6 observations taken five months apart suggests that the average flux density remains fairly stable. Similarly, in lower frequency ATCA data presented by \citet{li20}, the average flux densities in the 5.5 GHz and 9.0 GHz bands remained stable between observations taken in August 2016 and May 2017. Thus, we assume that the overall radio to mm SED gathered here is a representative of the average SED of the jet emission in \source.

\subsection{Spectral energy distribution}

\begin{figure*}
 \centering
 \includegraphics[width=\linewidth]{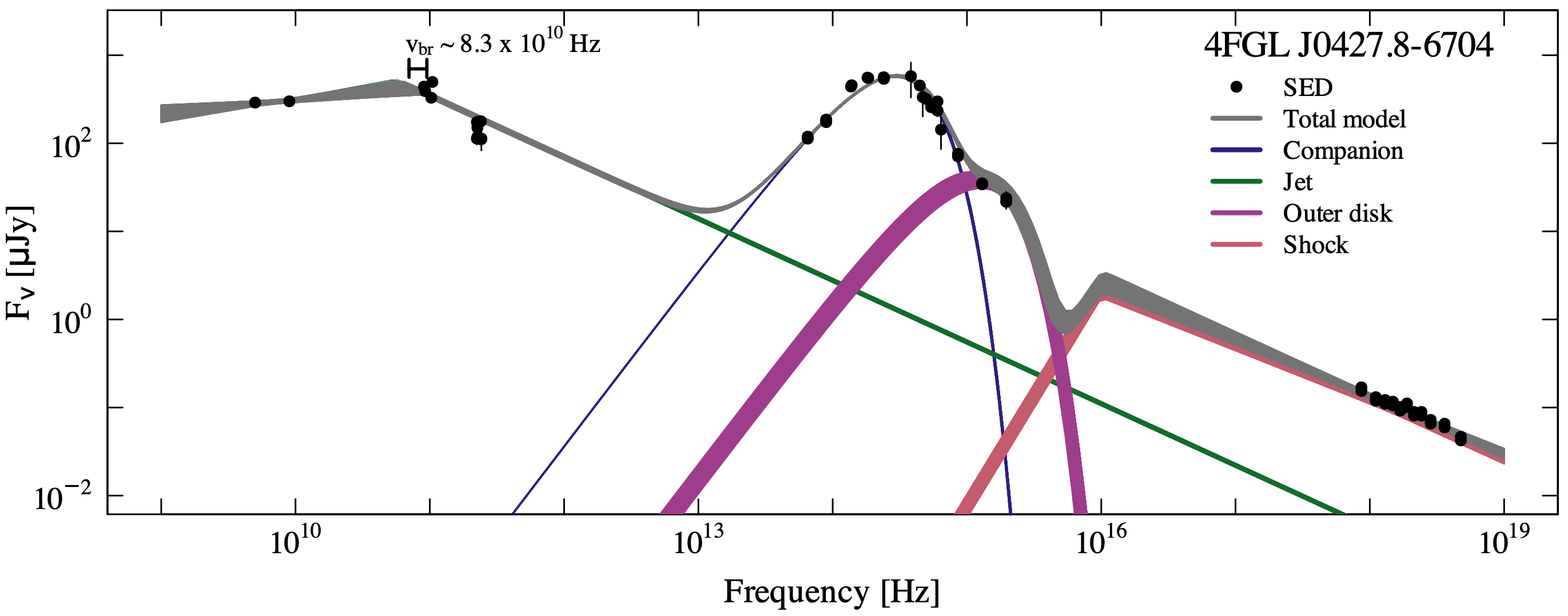}
 \includegraphics[width=\linewidth]{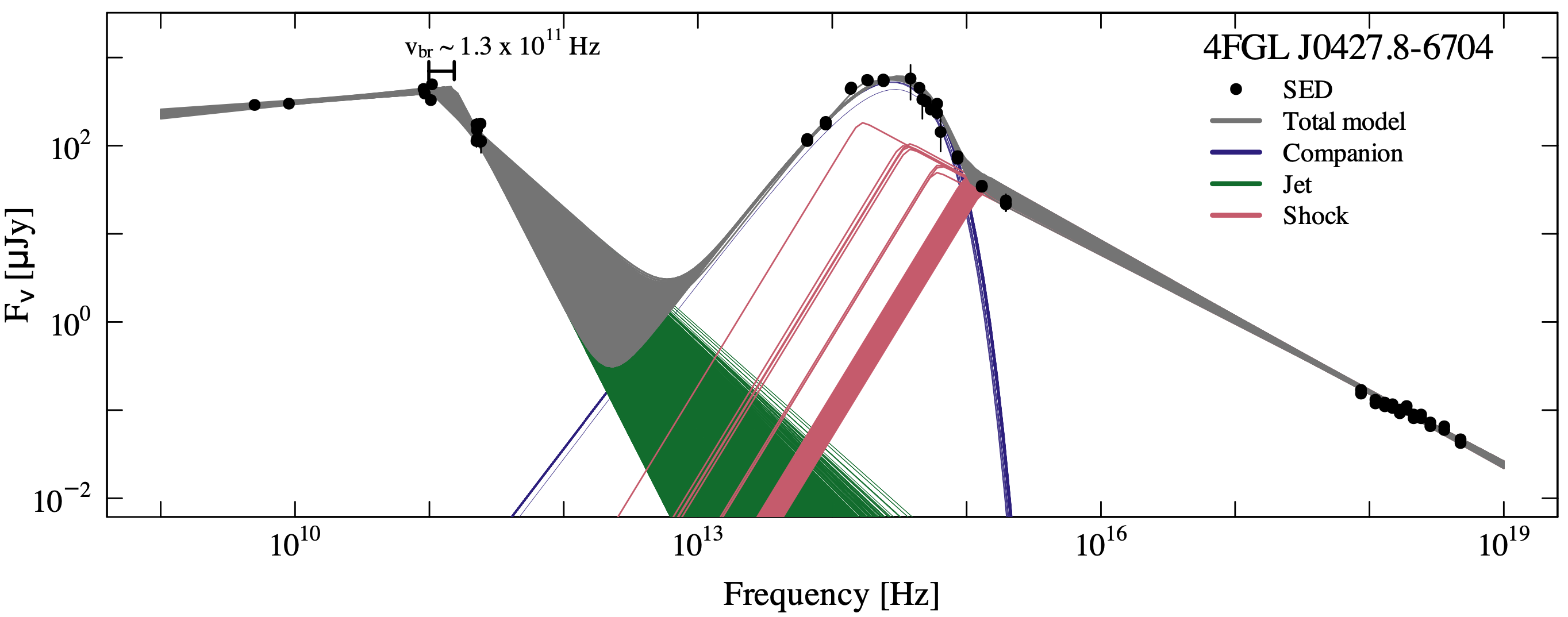}
 \caption{The SED of \source\/ is shown as black points. It is modeled using a phenomenological approach, incorporating two broken power law components to represent synchrotron emission from the jet and the termination shock at the inner edge of the accretion disk, along with a black body model for the companion star and the outer accretion disk (see text for further details, and Table 2 for the best-fit parameters). The downward-pointing arrows indicate upper limits from WISE (bands W3 and W4). The upper panel presents the model with the jet's optically thin spectral index fixed at $-$0.7 (Case I), while the lower panel shows the model where this parameter is left free (Case II). In addition, the ultraviolet data can be modeled in two ways: either by including an outer accretion disk component (used in the upper panel) or by extending the shock component into the ultraviolet/optical regime (used in the lower panel). Regardless of the model used, the jet break is constrained to $\sim$100 GHz (the 2$\sigma$ range is shown in the panels along with the median value).}
 \label{fig:sed}
\end{figure*} 

Fig. \ref{fig:sed} presents the collated SED with ALMA and NuSTAR data for \source\/ along with our phenomenological modeling described below. We note that the SED is not simultaneous and contains data taken in different years. However, the average radio/mm flux densities appear to remain relatively stable over time. 

As discussed above, ATCA 5/9 GHz average fluxes remained consistent between observations taken in August 2016 and May 2017 \citep{li20}. Our average fluxes from ALMA Band 3 observations (December 2017) align with the ATCA average flux densities, suggesting that the flat part of the radio SED below the break frequency does not vary significantly over the years. Similarly, our ALMA Band 6 average flux densities, obtained at different times (December 2017 and May 2018), are also comparable. As with lower frequencies, this indicates that the average flux remains relatively stable. 

The same applies to the X-ray emission, as demonstrated by \citet{strader16}, where the average X-ray flux showed no significant variation between the \textit{Swift} observations in October 2010 and the \textit{NuSTAR} observations in May 2016. We further checked the \textit{Swift} database\footnote{In particular, we used \textsc{swifttools} to build X-ray light curves of \source\/ available at \url{https://www.swift.ac.uk/user_objects/}.}, which contains data from 2010 to 2023, and found that the source flux is consistent with a weighted mean count rate of 0.013 cps, with a standard deviation of 0.006 cps, indicating that the X-ray flux remains fairly constant. 

For the infrared-to-ultraviolet data, we use the `upper envelope' for SED fitting, selecting the highest data points in each photometric band to ensure that the data correspond fully to the sub-luminous disk state. In addition, we apply corrections for interstellar extinction using $E(B-V) = 0.04$ \citep{schlafly11}, the dust extinction model by \citet{gordon23}, and a hydrogen column density of $N_{\rm H} = 7.7\times 10^{22}$ cm$^{-2}$ acquired from fitting only the X-ray data, using the abundances from \citet{wilms00}. 

The need for a high hydrogen column density in modeling the X-ray spectra of \source, as well as flux-dependent absorption, has been discussed in previous works and is attributed to a local component \citep{strader16,kennedy20,li20}, likely originating from the highly inclined accretion disk. The interstellar absorber accounts for $3.5 \times 10^{20}$ cm$^{-2}$ \citep{HI4pi}, which is consistent with the low interstellar extinction estimated towards the source. 

To perform the SED modeling, we use \textsc{ultranest} \citep{buchner21} that calculates the posterior probability distributions and the Bayesian evidence using the nested sampling Monte Carlo algorithm \textsc{MLFriends} \citep{buchner14,buchner19}. From the SED, we can clearly identify the compact jet spectrum, with a flat segment extending from radio to mm frequencies and a spectral cutoff around ALMA Band 3, followed by a steeper, optically thin spectrum. We model this jet spectrum phenomenologically using a broken power law model either fixing the optically thin spectral index at $\alpha_{\rm thin,jet}=-0.7$ (case I; corresponding to a non-thermal electron power law index of $p\sim2.5$), or letting it vary freely (case II), in which case we find $\alpha_{\rm thin,jet}=-2.1\pm0.4$ ($p\sim5$). We find similar jet break frequencies in either case; $v_{\rm br} \simeq 10^{10}$ Hz. In both scenarios, extrapolating this broken power law model to higher frequencies, we expect no significant contribution from jet emission in the optical-to-X-ray range.  

We simultaneously fit the companion star spectrum with a black body model and the X-ray data with a broken power law model (representing a synchrotron shock component), fixing the lower energy slope to $\alpha_{\rm thick,sh}=5/2$ appropriate for synchrotron self-absorption. The ultraviolet data clearly deviate from the black body model of the companion star, indicating an additional contribution from another component. This could originate either from an outer accretion disk -- though we note that no optical spectra of the source are available, leaving its exact contribution uncertain -- or from an extension of synchrotron shock emission to lower frequencies. Evidence of synchrotron shock emission in the optical has been observed in PSR~J1023$+$0038 in the high mode, where pulsed optical emission has been detected \citep{papitto19}. To account for both scenarios, we perform separate fits using these two different ultraviolet model components. An example of the fit using outer accretion disk component (approximated as a black body) is shown in Fig. 1 (upper panel), while the fit including the extension of the shock component is shown in the lower panel of Fig. 1. All model parameters for the four different scenarios (case I and II for the jet component, and disk and shock for the ultraviolet model component) are presented in Table \ref{tab:fit}.

Regardless of the chosen scenario, the resulting black body temperature is $T_{\rm bb} \approx 5100$ K, with a flux of $F_{\rm bb} \approx 2.8 \times 10^{-12}$ erg cm$^{-2}$ s$^{-1}$. The temperature is consistent with the value estimated from optical modeling in \citet{kennedy20}, where the companion temperature was estimated to be 5300$\pm$600 K based on light curve modeling. When including an outer disk component, the resulting black body temperature is $T_{\rm bb} \approx 21000$ K, with a flux of $F_{\rm bb} \approx 2.4 \times 10^{-9}$ erg cm$^{-2}$ s$^{-1}$. In this case, we fix the break frequency of the shock model to $\nu_{\rm br, shock}=10^{16}$ Hz. However, if we allow the shock break frequency to extend to lower frequencies, the outer accretion disk component is no longer required by the fit, and the spectral index of the shock power law component is constrained to $-0.82\pm0.01$, consistent with non-thermal synchrotron emission.

\begin{table*}
    \caption{Best-fit parameters from our SED modeling of \source. The SED was modeled using a phenomenological approach including either two broken power law components and one black body to represent synchrotron emission from the jet and shock, as well as thermal emission from the companion star, or adding an additional black body component to the above model to account for the outer disk emission.} 
    \centering
    \begin{tabular}{c|ccccccc}   
    \hline
    Component & Model$^{a}$ & Case$^{b}$ & UV model$^{c}$ & \multicolumn{4}{c}{Parameters} \\
    \hline
     \multirow{3}{*}{Jet} & \multirow{3}{*}{BPL} & & & $\alpha_{\rm thick, jet}$ & $\alpha_{\rm thin, jet}$ & $\nu_{\rm br, jet}$ & S$_{\nu_{\rm br}, {\rm jet}}$  \\
         &   & I & \multirow{2}{*}{Shock} & 0.12$\pm$0.03 & $-$0.7 (fixed) & (8.3$\pm$0.6)$\times10^{10}$ Hz & 0.40$\pm$0.02 mJy\\
         &   & II & & 0.131$\pm$0.009 & $-$2.1$\pm$0.4 & (12.6$\pm$1.3)$\times10^{10}$ Hz & 0.43$\pm$0.01 mJy\\    
         &   & I & \multirow{2}{*}{Disk} & 0.12$\pm$0.03 & $-$0.7 (fixed) & (8.3$\pm$0.6)$\times10^{10}$ Hz & 0.40$\pm$0.03 mJy\\
         &   & II & & 0.131$\pm$0.009 & $-$2.1$\pm$0.4 & (12.6$\pm$1.4)$\times10^{10}$ Hz & 0.43$\pm$0.01 mJy\\          
     \multirow{3}{*}{Companion} & \multirow{3}{*}{BB} & & & $T_{\rm bb}$ & $F_{\rm bol}$ \\
         &   & I & \multirow{2}{*}{Shock} & 5129$\pm$30 K & (2.9$\pm$0.1)$\times10^{-12}$ erg s$^{-1}$ cm$^{-2}$ \\
         &   & II & & 5100$\pm$32 K & (2.9$\pm$0.1)$\times10^{-12}$ erg s$^{-1}$ cm$^{-2}$ \\      
         &   & I & \multirow{2}{*}{Disk} & 5067$\pm$28 K & (2.8$\pm$0.1)$\times10^{-12}$ erg s$^{-1}$ cm$^{-2}$ \\
         &   & II & & 5038$\pm$28 K & (2.8$\pm$0.1)$\times10^{-12}$ erg s$^{-1}$ cm$^{-2}$ \\            
     \multirow{3}{*}{Shock} & \multirow{3}{*}{BPL} & & & $\alpha_{\rm thick, sh}$ & $\alpha_{\rm thin, sh}$ & $\nu_{\rm br, sh}$ & S$_{\nu_{\rm br}, {\rm sh}}$  \\
         &   & I & \multirow{2}{*}{Shock} & 2.5 (fixed) & $-$0.83$\pm$0.01 & (1.22$\pm$0.08)$\times10^{15}$ Hz & 0.037$\pm$0.006 mJy\\
         &   & II & & 2.5 (fixed) & $-$0.82$\pm$0.01 & (1.21$\pm$0.07)$\times10^{15}$ Hz & 0.038$\pm$0.003 mJy\\   
         &   & I & \multirow{2}{*}{Disk} & 2.5 (fixed) & $-$0.66$\pm$0.02 & 10$^{16}$ Hz (fixed) & 0.0026$\pm$0.0003 mJy\\
         &   & II & & 2.5 (fixed) & $-$0.66$\pm$0.02 & 10$^{16}$ Hz (fixed) & 0.0027$\pm$0.0003 mJy\\   
     \multirow{3}{*}{Outer disk} & \multirow{3}{*}{BB} & & & $T_{\rm bb}$ & $F_{\rm bol}$ \\     
         &   & I & \multirow{2}{*}{Disk} & 21138$\pm$683 K & (2.3$\pm$0.4)$\times10^{-9}$ erg s$^{-1}$ cm$^{-2}$ \\
         &   & II & & 21043$\pm$720 K & (2.4$\pm$0.4)$\times10^{-9}$ erg s$^{-1}$ cm$^{-2}$ \\          
    \hline
    \multicolumn{8}{l}{$^{a}$ BPL: Broken power law, BB: Black body.} \\
    \multicolumn{8}{l}{$^{b}$ I: The power law index above break frequency is fixed to $-0.7$, II: The power law index above break frequency is left as a free parameter.} \\
    \multicolumn{8}{l}{$^{c}$ Fitting strategy for modeling the SED at UV regime. Shock: allowing the extension of the shock power law break frequency to UV/optical regime,}\\ 
    \multicolumn{8}{l}{\hspace{0.2cm}Disk: adding additional black body component to the model and fixing the shock power law break frequency to $10^{16}$ Hz.} 
    \end{tabular}
    \label{tab:fit}
\end{table*}

\section{Discussion} \label{sec:discuss}

\begin{figure}
 \centering
 \includegraphics[width=\linewidth]{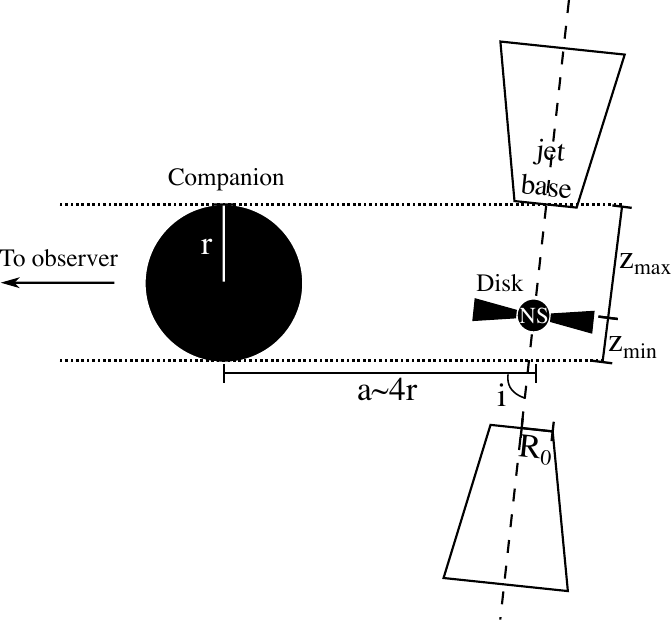}
 \caption{The system geometry at eclipse. The neutron star is not in scale. The high energy emission arising close to the neutron star is eclipsed. However, the flux from the jet base does not show eclipses and therefore has to arise further out than $z_{\rm max}$.}
 \label{fig:cartoon}
\end{figure} 

Due to the system's high orbital inclination, we can place constraints on the location of the mm emission based on the absence of observed radio eclipses. Since the break occurs near ALMA Band 3, these constraints can be attributed to the jet base or the first acceleration region, where the radio emission is expected to transition from the optically thick to thin regime. In addition, we derive physical properties from the conical jet scenario \citep{blandford79} and jet power from minimum energy requirements for synchrotron radiation \citep{burbidge56}. 

\subsection{Jet size constraints}

The jet is eclipsed by the companion star between the following bounds:
\begin{equation}
    a \mathrm{cot}(i) - r \mathrm{csc}(i) < z < a \mathrm{cot}(i) + r \mathrm{csc}(i)
\end{equation}

\noindent where $a$ is the orbital separation, $i$ is the orbital inclination, $r$ is radius of the companion, and $z$ is the jet height \citep[see Fig. \ref{fig:cartoon};][]{maccarone20}. Assuming the neutron star and companion masses are $M_{\mathrm{NS}} = 1.4 M_{\odot}$ and $M_{\mathrm{C}} = 0.3 M_{\odot}$, respectively, and that the companion fills its Roche Lobe, using an orbital inclination of $i=84\pm3^{\circ}$ \citep{kennedy20} we obtain $(-2\pm1) \times 10^{10}$ cm $< z <$ $(7\pm1) \times 10^{10}$ cm (negative values correspond to the counter jet). Since there is no evidence of eclipses in the radio/mm emission, the jet base must be located beyond $z_{0} > 7 \times 10^{10}$ cm. This distance is significantly larger than jet base estimates from black hole XRBs, which measure around $3 \times 10^{9}$ cm, based on X-ray to optical time lags \citep{gandhi17}. However, their jet breaks occur in the near-infrared region indicating smaller emitting areas and likely closer locations with respect to the compact object. 

Assuming that the cylindrical region at the jet base (the particle acceleration site) consists of synchrotron-emitting non-thermal electrons following a power law distribution, we can use the jet break frequency and flux density to estimate its size \citep{chaty11}. We follow the formalism outlined in \citet[][their Appendix C]{russell20}. The relation is given by: 

\begin{eqnarray}
R_{0} \propto & \nu_{\rm br}^{-1} (S_{\nu_{\rm br}}/h)^{(p+6)/(2p+13)} d^{(2p+12)/(2p+13)} \xi^{-1/(2p+13)}
\\
R_{0} \approx & 2 \times 10^{10} \Big( \frac{8.3\times10^{10} \, \mathrm{Hz}}{\nu_{b}} \Big) \Big( \frac{S_{\nu_{br}}}{0.2 \, \mathrm{mJy}} \Big)^{17/36} \Big( \frac{d}{2.5 \, \mathrm{kpc}} \Big)^{17/18} \, \mathrm{cm}
\\
R_{0} \approx & 1 \times 10^{10} \Big( \frac{12.6\times10^{10} \, \mathrm{Hz}}{\nu_{b}} \Big) \Big( \frac{S_{\nu_{br}}}{0.2 \, \mathrm{mJy}} \Big)^{10/21} \Big( \frac{d}{2.5 \, \mathrm{kpc}} \Big)^{20/21} \, \mathrm{cm}
\end{eqnarray}

\noindent where, in Eq. 3, we used $p=2.5$ (case I), and in Eq. 4, $p=5$ (case II). The equation is not highly sensitive to the value of $p$. In addition, we assume equipartition between the energy of the non-thermal electrons and magnetic energy density, $\xi=(u_{e}/3)/(B^2/8\pi)=1$, and no significant proton contribution. While these assumptions are common, we note that there is no particular physical justification for these in our case. However, our constraints on jet size (Eq. 2) are not particularly sensitive to the equipartition fraction. The maximum and minimum Lorentz factors for the electron power law distribution are assumed to be $\gamma_{\mathrm{max}} \gg \gamma_{\mathrm{min}}=1$. The emitting region is homogeneous, with its height equal to the cylinder's radius, $R_{0}=hH_{0}$, where $h=1$. The synchrotron radiation is averaged over an isotropic distribution of the electron pitch angles. 

This formulation corresponds to a one-zone scenario, so we divided the jet flux density by a factor of two to account for counter jet emission, neglecting Doppler boosting given the high inclination and likely slow jet speed \citep[the only neutron star compact jet velocity measurement gives $v\sim0.4c$;][]{russell24}. The resulting radii ($R_{0}=(1-2)\times10^{10}$ cm) are consistent with the light crossing time limit derived above ($R_{0}<6\times10^{10}$ cm). 

Since the jet is not resolved, the highest resolution ALMA Band 6 observations place an upper limit on the size of the jet, which is 0.1 arcsec (major axis FWHM), that at the distance of 2.5 kpc corresponds to 3.7$\times 10^{15}$ cm.  

To estimate the opening angle of the jet, we use the relation tan$(\phi/2) = R_{0}/z_{0}$, where $z_{0}$ is the distance from the neutron star to the jet base. This relation assumes a conical jet with a constant opening angle $\phi$. Assuming $R_{0} = 2 \times 10^{10}$ cm and $z_{0} > 7 \times 10^{10}$ cm, we obtain an upper limit for the opening angle of $\phi <$ 32$^{\circ}$. This limit is not very restrictive, as typical opening angles for XRB jets measure $<10^{\circ}$ \citep[e.g.,][]{millerjones06,russell19,tetarenko19b,espinasse20,carotenuto21,chauhan21,tetarenko21,wood24}.

\subsection{Magnetic field strength at the jet base}

The jet break frequency and flux density can also provide an estimate of the magnetic field strength at the first acceleration region \citep{chaty11}. Following the formalism of \citet{russell20}, the magnetic field strength at the jet base is given by:

\begin{eqnarray}
B_{0} \propto & \nu_{b} S_{\nu_{b}}^{-2/(2p+13)} (\xi d)^{-4/(2p+13)},
\\
B_{0} \approx & 118 \Big( \frac{\nu_{b}}{8.3\times10^{10} \, \mathrm{Hz}} \Big) \Big( \frac{S_{\nu_{br}}}{0.2 \, \mathrm{mJy}} \Big)^{-1/9} \Big( \frac{d}{2.5 \, \mathrm{kpc}} \Big)^{-2/9} \, \mathrm{G},
\\
B_{0} \approx & 468 \Big( \frac{\nu_{b}}{12.6\times10^{10} \, \mathrm{Hz}} \Big) \Big( \frac{S_{\nu_{br}}}{0.2 \, \mathrm{mJy}} \Big)^{-2/21} \Big( \frac{d}{2.5 \, \mathrm{kpc}} \Big)^{-4/21} \, \mathrm{G},
\end{eqnarray}

\noindent where, in Eq. 6, we used $p=2.5$ (case I), and in Eq. 7, $p=5$ (case II). Since the magnetic field strength at the jet base is directly proportional to the jet break frequency, most hard-state XRB jets exhibit relatively high values of magnetic field strength, with $B_{0} \sim 10^4$ G \citep{chaty11,russell14,russell20,echiburu24}.

The low magnetic field strength of \source\/ implies that the synchrotron cooling time at the jet base is relatively long for electrons with estimated Lorentz factors of $\gamma = \sqrt{\nu / 2.8\times10^6 B_{0}} \approx 11-17$, where $\nu=97.5$ GHz:

\begin{equation}
    t_{\rm syn} = \frac{6 \pi m_{\rm e} c}{\sigma_{\rm t} B_0^2 \gamma} \sim 10^3 \, \mathrm{s},
\end{equation}

\noindent where $m_{\rm e}$ is the electron mass, and $\sigma_{\rm t}$ is the Thomson cross-section. In contrast, the adiabatic timescale at the jet base is much shorter: $t_{\rm ad} \gtrsim R_{0}/\beta_{\rm exp} c \approx 0.6$ s, where $\beta_{\rm exp}$ is the expansion speed of the ejected material, assumed to be the relativistic sound speed ($\beta_{\rm exp}=1/\sqrt{3}$). The adiabatic timescale is consistent with observed second-scale flares, suggesting that the mm flaring behavior in \source\/ is driven by continuously ejected blobs or shocks that cool adiabatically within the jet.

\subsection{Jet power}

We estimate the minimum power required to produce the fast synchrotron flares observed in the ALMA Band 3 light curve using the formulation presented in \citet{longair11} and \citet{fender06}:

\begin{equation}
P_{\rm min} \sim (0.5-3.5)\times10^{33} \Big( \frac{\Delta t}{\rm s} \Big)^{2/7} \Big( \frac{\nu}{\rm GHz} \Big)^{2/7} \Big( \frac{\Delta S_{\nu}}{\rm mJy} \Big)^{4/7} \Big( \frac{d}{\rm kpc} \Big)^{8/7} \, {\rm erg} \, {\rm s}^{-1},
\end{equation}

\noindent where the range corresponds to Case II (lower value) and Case I (higher value). For the 2-s flare with an amplitude of 1.3 mJy, we find $P_{\rm min} \sim (0.8-5.2)\times 10^{34}$ erg s$^{-1}$. For the longer duration flare observed at the end of the observation ($\Delta$t $\sim$ 625 s, $\Delta$S$_{\nu}$ $\sim$ 0.5 mJy), the power is $P_{\rm min} \sim (2.3-15.6)\times 10^{34}$ erg s$^{-1}$. Therefore, the average jet power would be at least on the order of $P_{\rm jet} \sim 10^{35}$ erg s$^{-1}$. This is one to two orders of magnitude higher than the X-ray luminosity ($P_{\rm min}/L_{\rm X} \sim 10-100$), notably different from other black hole or neutron star XRBs, which show $P_{\rm min}/L_{\rm X} = 0.1-1$ \citep{gallo05,russell13b,tetarenko21,russell24}. This disparity could be due to the inefficient X-ray emission mechanism in tMSPs, which produces much lower X-ray luminosities compared to other XRB systems. In particular, X-ray emission in tMSPs is not directly linked to accretion, and it has been suggested to arise from the shock between the pulsar wind and matter in the disk \citep{papitto19,veledina19}. However, X-ray luminosity could still partially track the mass accretion rate, as increased material accumulation at the disk truncation radius creates a thicker disk and a larger surface for the pulsar wind to interact with. Alternatively, the relatively high jet power in \source\/ suggests that the jet may be drawing power from a source other than accretion, as we discuss in more detail in Section \ref{sec:discuss:XRB}.

The corresponding equipartition magnetic field strength for the above minimum jet power is:

\begin{equation}
B_{\rm eq} \sim (7-18) \Big( \frac{\Delta t}{\rm s} \Big)^{-6/7} \Big( \frac{\nu}{\rm GHz} \Big)^{1/7} \Big( \frac{\Delta S_{\nu}}{\rm mJy} \Big)^{2/7} \Big( \frac{d}{\rm kpc} \Big)^{4/7} \, {\rm G},
\end{equation}

\noindent where the range again corresponds to Case II (lower value) and Case I (higher value). For the 2-s flare, $B_{\rm eq}$ is between 14 and 35 G, and the slower 625-s flare it is between 0.1 and 0.2 G. These values are 1--3 orders of magnitude lower than the estimates for the jet base provided in Section 4.2. This discrepancy could arise from the jet base not being in equipartition, the emitting volume being smaller than inferred from the light-crossing time, or the flare timescale being shorter than what our data can resolve.

\subsection{Jet radiative luminosity}

Using the SED fits, we can estimate the radiative luminosity, $L_{\rm j}$, of the jets. We integrate the model luminosities from $10^{9}$ Hz up to $10^{14}$ Hz. This upper frequency corresponds roughly to the value of the synchrotron cooling break observed in black hole XRB MAXI J1836-194, when the jet break was located at $\sim10^{11}$ Hz \citep{russell14}. For Case II, with a steep slope after the break, the value for the cooling break frequency does not significantly affect the total radiative luminosity, which is approximately $10^{30}$ erg s$^{-1}$. In Case I, the radiative luminosity (up to $10^{14}$ Hz) is $6\times10^{30}$ erg s$^{-1}$. In both cases, the radiative efficiency is low compared to the minimum power derived above, $L_{\rm j}/P_{\rm min} \lesssim$0.01\%. It is at least two orders of magnitude lower than the canonical estimates for XRBs in the hard state \citep{blandford79,fender00,fender01,gallo05,russell07}, mainly arising from the three orders of magnitude lower jet break frequency. 

\subsection{Comparison to other XRBs: luminosities} \label{sec:discuss:XRB}

\begin{figure}
 \centering
 \includegraphics[width=\linewidth]{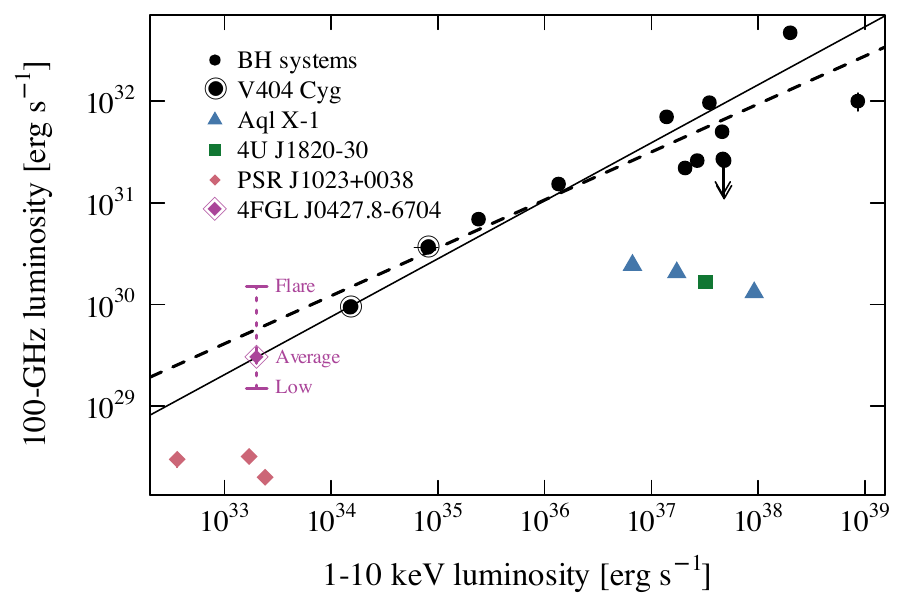}
 \caption{100 GHz luminosity as a function of 1--10 keV X-ray luminosity (adapted from \citealt{baglio23}). Here, we add \source\/ to their dataset. Due to the high mm emission with respect to the other tMSP PSR J1023+0038, \source\/ seems to line up with the best-fit relation for the black hole XRBs (dashed line). Solid line show theoretical relation of $L_{\rm J} \propto \dot{M}^{4/7}$ from \citet{parfrey16} for jets powered by enhanced stellar open flux (scaled to match the average mm luminosity of \source).}
 \label{fig:corr}
\end{figure} 

Following \citet{baglio23}, we plot the mm and X-ray luminosities, $L_{\rm mm}$ and $L_{\rm X}$, respectively, in Fig. \ref{fig:corr}. Interestingly, \source\/ aligns with black hole XRBs, following the relation $L_{\rm mm} \propto L_{\rm X}^{0.48\pm0.07}$, rather than with neutron star XRBs, which exhibit lower $L_{\rm mm}$ for a given $L_{\rm X}$. Given the flat jet spectrum, a similar finding was already discussed in \citet{li20} considering the $L_{\rm 5 GHz}/L_{X}$ plane. 

This alignment is surprising for two reasons. First, if the X-ray luminosity originates from the synchrotron emission at the shock formed between the pulsar wind and the truncated inner edge of the accretion disk \citep{papitto19,veledina19}, then the X-ray luminosity is driven by the pulsar spin-down luminosity ($\dot{E}\sim10^{34-35}$ erg s$^{-1}$ generally for compact binary millisecond pulsars; e.g., \citealt{koljonen23}) rather than accretion. In this case, the location of the source in the $L_{\rm mm}/L_{X}$ plane alongside accreting systems would be purely coincidental. Secondly, if the X-ray luminosity results from a low accretion rate onto the neutron star, then its radio/mm luminosity is much higher than that of other neutron star XRBs given its low X-ray luminosity ($L_{X} \approx 10^{-5} L_{\rm Edd}$).

Overall, hard state black hole XRBs approximately follow a relation between the radio and X-ray luminosities, $L_{\rm R} \propto L_{\rm X}^{0.6}$ \citep[e.g.,][]{hannikainen98,corbel03,gallo03,gallo18,koljonen19}, similar to that of $L_{\rm mm}$/$L_{\rm X}$ plane shown in Fig. \ref{fig:corr}, which is not surprising considering the flat jet spectrum. This relation can be quantitatively explained by the Blandford-Znajek process \citep{jiang24}, where magnetic field generated at the outer disk is advected inwards by the inner hot accretion flow and substantially enhanced close to the compact object leading to increased jet power and accretion luminosity. Given the location of \source\/ in the $L_{\rm mm}/L_{X}$ plane, a similar mechanism to the Blandford-Znajek process may be at work in \source. Indeed, a comparable spin-powered scenario can be applied to neutron star systems, where the jet could be powered by the opening of the neutron star's magnetic field lines due to accreting matter \citep{parfrey16,parfrey17}. This model also predicts a similar relationship between the accretion rate and jet power, with $L_{\rm J} \propto \dot{M}^{4/7}$, as observed in XRBs.

If all tMSPs adhere to the same jet-launching mechanism, the order-of-magnitude difference in 100 GHz luminosity between \source\/ and PSR~J1023$+$0038 must be explained. In the spin-powered scenario by \citet{parfrey16}, in addition to the accretion rate, the jet power depends on the magnetic dipole moment, $\mu$, neutron star mass, $M_{\rm NS}$, spin frequency, $\nu$, and two parameters related to the efficiency of opening the magnetic field lines, $\zeta$, and disk trunction, $\xi$, with values close to unity:

\begin{multline}
    L_{\rm jet} = 1.6\times10^{36} \Big( \frac{\zeta}{\xi} \Big)^2 \Big( \frac{\nu}{\rm 500 \, Hz} \Big)^2 \Big( \frac{\mu}{\rm 10^{26} \, G \, cm^3} \Big)^{6/7} \\ \times \Big( \frac{M_{\rm NS}}{1.4 M_{\odot}} \Big)^{6/7} \Big( \frac{\dot{M}}{\dot{M}_{\rm Edd,\odot}} \Big)^{4/7} \, \mathrm{erg} \, \mathrm{s^{-1}}. 
\end{multline}

\noindent Given that the neutron star mass and X-ray luminosity are similar between \source\/ and PSR~J1023$+$0038, the difference in jet power (and luminosity) may be attributed to a combination of different magnetic dipole moments and/or spin frequency. The magnetic dipole moment for PSR~J1023$+$0038 can be estimated using a magnetic field strength of $B=9.7\times10^7$ G \citep{archibald13} and a neutron star radius of $R=12$ km, yielding $\mu=BR^3=1.7\times10^{26}$ G cm$^{-3}$. A tenfold increase in jet power would therefore require $\mu=BR^3=1.8\times10^{27}$ G cm$^{-3}$ ($B=10^9$ G for $R=12$ km) for \source, assuming the other parameters are similar to those of PSR~J1023$+$0038 (including spin frequency). The higher magnetic moment could also explain why \source\/ is dominated by flares and lacks the typical high/low mode bimodality. A stronger magnetic field may prevent the accretion of matter inside the light cylinder, thereby inhibiting the system from entering the low mode \citep{veledina19}. Simultaneous X-ray and radio observations of PSR~J1023$+$0038 and the tMSP candidate CXOU~J110926.4$-$650224 in flaring mode have shown that X-ray flares precede radio flares \citep{bogdanov18,cotizelati21}. This behavior is thought to result from magnetic reconnection events within the disk or at the turbulent termination shock, potentially launching plasmoids that sporadically increase the mass-loading of the jet resulting in synchrotron blobs that become optically thin at the jet base \citep{campana19,cotizelati21}.

\subsection{Comparison to other XRBs: jet break frequencies}

In most XRB outbursts, during the hard state ($10^{-3}<L_{\rm X}/L_{\rm Edd}<0.5$), the compact jet dominates the radio-mm emission, and the jet break typically appears in the optical-infrared range for both black hole \citep[see, e.g.][]{russell13, koljonen15} and neutron star XRBs \citep{migliari10, diaztrigo17}. Instances of lower frequency jet breaks in XRBs are observed during state transitions from the hard to soft state occurring (typically) during the outburst rise, where the jet break shifts rapidly (within weeks to hours) from higher frequencies ($10^{13-14}$ Hz) down to $10^{10}$ Hz \citep{russell14,russell20}. These state changes are connected to changes in the accretion flow, which accretes at a high rate ($\sim0.1-0.3 \, \dot{m}_{\rm Edd}$; \citealt{gierlinski06}), resulting in jet quenching and transient jet ejections that likely affect conditions in the first acceleration region \citep[e.g.,][]{koljonen15,russell20}. During the outburst decay, the source changes back to the hard state at a lower accretion rate ($\sim0.01-0.04 \, \dot{m}_{\rm Edd}$; \citealt{maccarone03}), the jet emission recovers, and the jet break frequency gradually increases from radio/mm wavelengths back to near-infrared \citep{echiburu24}. From Eq. 2, since the radius of the acceleration region is directly proportional to the jet break frequency, this suggests that the first acceleration region becomes smaller and progressively moves closer to the compact object.

Among the black hole XRBs, the closest comparison to \source\/ in terms of X-ray and radio luminosity is V404 Cyg during the decay phase of its 2015 outburst \citep{tetarenko19}. While its radio and X-ray luminosities are quite similar to those of \source\/ (see Fig. \ref{fig:corr}), the jet break in V404 Cyg was observed at $\sim10^{14}$ Hz \citep{tetarenko19}, about three orders of magnitude higher than in \source. This suggests that, although both sources are best described by a compact, partially self-absorbed jet, the properties of the first acceleration region differ significantly between these systems.

Jet break measurements in neutron star XRBs are scarce. For example, 4U 1728$-$34 and 4U 0614$+$091, which are atoll-type, low-magnetic field, persistently accreting neutron star XRBs, have jet breaks measured at $\sim6\times10^{13}$ Hz \citep{diaztrigo17} and $[1-4]\times10^{13}$ Hz \citep{migliari10}, respectively. In 1RXS~J180408.9$-$342058, a jet break has not been observed, but the SED of the source exhibits a flat spectrum in the near-infrared, likely indicating the presence of a (transient) jet with a jet break above $10^{15}$ Hz \citep{baglio16}. In Aql X-1, a transient neutron star low-mass XRB, which shows jet break evolution similar to that of black hole XRBs, though beginning at a lower jet break frequency (10$^{11}$ GHz at the outburst onset), shifting below 5 GHz as the outburst evolves to the soft state, and recovering to 10$^{11}$ GHz in the outburst decay \citep{diaztrigo17}. Due to the stable average radio and mm luminosities in \source, along with the likely very low accretion rate \citep[with accretion rates in tMSPs estimated at $\dot{M}\sim 10^{-5}\dot{M}_{\rm Edd}$;][]{linares14}, the jet in \source\/ appears most similar to the compact jet observed in the hard state of Aql X-1, with a comparable jet break frequency and similar jet luminosity. However, Aql X-1 is significantly more X-ray luminous than \source, by at least three orders of magnitude ($L_{\rm X}/L_{\rm Edd}=0.04-0.1$ in the hard state), suggesting that the jet break frequency and the properties of the jet base are not directly related to the mass accretion rate. In addition, the origin of X-ray emission likely differs between these sources. In tMSPs, X-ray emission is thought to originate from synchrotron radiation produced by the shock between inflowing matter and the pulsar wind \citep{papitto19,veledina19}. In contrast, in low-mass XRBs during the hard state, X-ray emission primarily arises from inverse Compton scattering of accretion disk photons by a hot plasma near the compact object. Nonetheless, it is intriguing to consider what unites Aql X-1 and \source\/ in their low-frequency jet breaks. Notably, coherent millisecond X-ray pulsations have been observed in Aql X-1, indicating a relatively strong neutron star magnetic field ($10^8 \, \mathrm{G} < B < 2\times10^9$ G) capable of channeling accretion disk material to the magnetic poles \citep{casella08}. This upper limit on magnetic field strength overlaps with our estimate for \source, suggesting that neutron star properties in Aql X-1 may be similar to those in tMSPs, which could influence the characteristics of the first acceleration region.

\subsection{SEDs for other tMSP systems}

We also compiled the SEDs from other tMSP systems and candidates, including PSR J1023$+$0038, 4FGL J1544.5$-$1126, and XSS J12270$-$4859 (see Appendix \ref{appendix}). However, the jet breaks in these systems are not well constrained by the available data. We note that while the jet break in PSR~J1023$+$0038 has been estimated to occur around 2.5$\times 10^{13}$ Hz \citep{baglio23}, there is insufficient data in the mm-to-infrared range to definitely constrain its location (we estimate a range from $10^{11}$ Hz to $10^{13}$ Hz; see Appendix A), unlike for \source.The most promising indication of a high-frequency jet break can be seen for XSS J12270$-$4859, as the source has detections in the WISE W3-band ($2.6\times10^{13}$ Hz). These detections are difficult to explain arising from the companion star that has a non-irradiated base temperature of 5500 K \citep{riverasandoval18,stringer21}. If the WISE data arises from the compact jet, we restrict its jet break to $\sim10^{14}$ Hz, a similar value than for hard state XRBs. Unfortunately, XSS J12270$-$4859 is currently in a pulsar state, and a transition to a sub-luminous disk state would be needed to confirm the nature of the mid-infrared emission. Overall, further observations with millimeter and mid-infrared facilities would be required to better constrain the jet spectra of other tMSPs and candidates.

\section{Conclusion} \label{sec:conclude}

In this paper, we compiled the SEDs of tMSPs to study their jet properties. Specifically, we analyzed ALMA data from \source\/ and derived the frequency ($v_{\rm br}\simeq 10^{11}$ Hz) and flux density ($S_{\rm br}\simeq0.4$ mJy) of its jet break, providing direct estimates of the jet base properties. The millimeter emission does not exhibit eclipses, unlike in optical, X-ray, and $\gamma$-ray wavelengths, suggesting that the jet base emission is located farther out in the jet ($z_{0}>7\times10^{10}$ cm) compared to estimates for other XRBs. However, the millimeter luminosity of \source\/ is consistent with the millimeter/X-ray scaling relation observed in black hole XRBs, and is an order of magnitude brighter than PSR J1023$+$0038. In addition, the estimated jet power agrees with those from other neutron star systems. A similarly stable, low-frequency jet break has only been inferred in one other system, Aql X-1, which may indicate similarities in the jet bases of these two sources, potentially influenced by the relatively strong magnetic fields of the neutron stars. Overall, tMSP jets provide a unique window into understanding low-mass accretion rate jets and the impact of strong magnetic fields on their properties.    

\section*{Acknowledgements}

We thank Thomas Russell and Rapha\"el Mignon-Risse for discussions on deriving the jet properties and jet launching mechanisms, respectively. 

This project has received funding from the European Research Council (ERC) under the European Union’s Horizon 2020 research and innovation programme (grant agreement No. 101002352, PI: M. Linares).

JCAM-J was the recipient of an Australian Research Council Future Fellowship (FT140101082).

This paper makes use of the following ALMA data: ADS/JAO.ALMA\#2017.A.00035.S and ADS/JAO.ALMA\#2017.1.01690.S. ALMA is a partnership of ESO (representing its member states), NSF (USA) and NINS (Japan), together with NRC (Canada), MOST and ASIAA (Taiwan), and KASI (Republic of Korea), in cooperation with the Republic of Chile. The Joint ALMA Observatory is operated by ESO, AUI/NRAO and NAOJ.

This research has made use of data and/or software provided by the High Energy Astrophysics Science Archive Research Center (HEASARC), which is a service of the Astrophysics Science Division at NASA/GSFC.

This research has made use of the VizieR catalogue access tool, CDS,
Strasbourg, France \citep{10.26093/cds/vizier}. The original description of the VizieR service was published in \citet{ochsenbein00}.

\section*{Data Availability}

All data is available at their respective observatory archives.



\bibliographystyle{mnras}
\bibliography{references} 




\appendix

\section{Spectral energy distributions of \lowercase{t}MSPs and candidates} \label{appendix}

\subsection{XSS J12270$-$4859}

\begin{figure*}
 \centering
 \includegraphics[width=\linewidth]{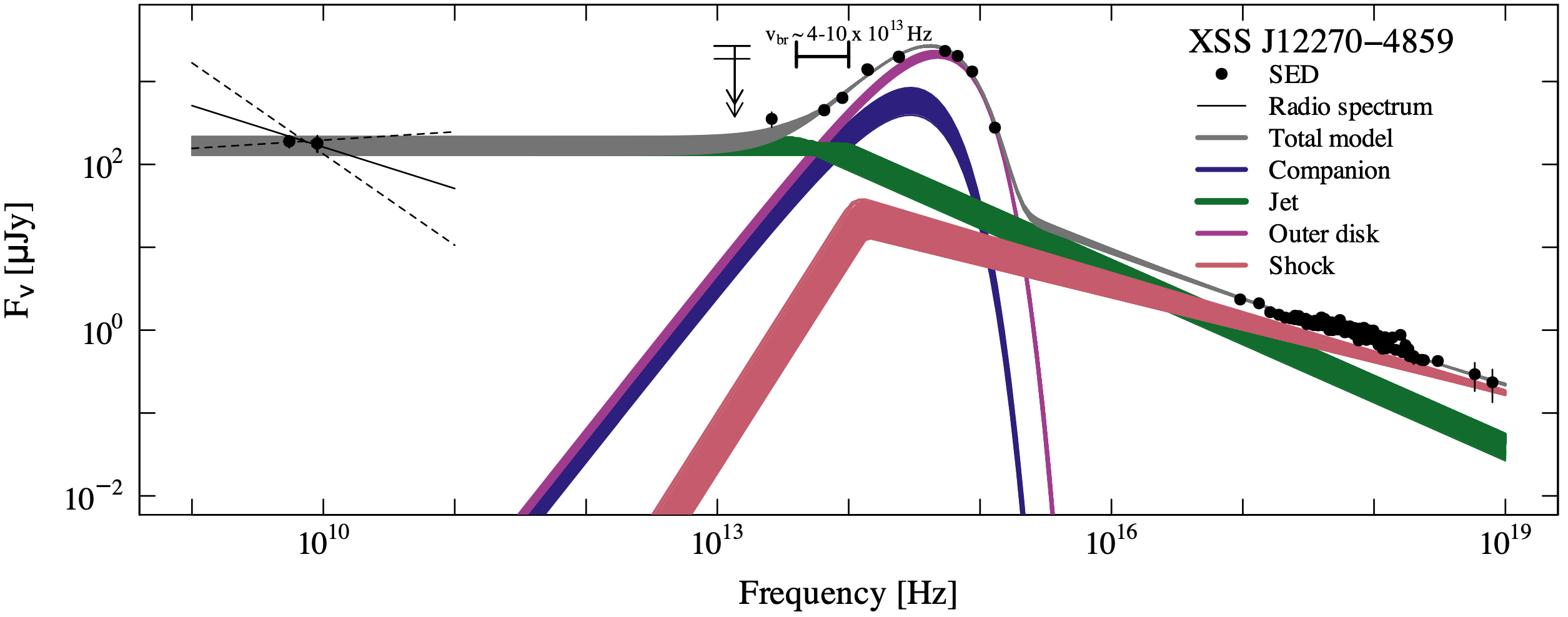}
 \caption{The SED of XSS J12270$-$4859 in the sub-luminous disk state is shown as black points. We fit the SED using the same model as for \source, as discussed in Section 3.3, that has the additional black body component representing emission from an outer accretion disk. The downward-pointing arrow indicate an upper limit from WISE (band W4). The solid and dashed black lines indicate the possible radio spectra according to the data. However, in the modeling, we fix the jet spectrum to $F_{\rm jet}\propto \nu^{0}$.}
 \label{fig:sed_J1227}
\end{figure*} 

XSS J12270$-$4859 was initially classified as a low-mass XRB accreting from a low-mass companion star \citep{demartino10}. However, in 2013, a significant transition occurred when radio pulsations were detected, identifying XSS J12270$-$4859 as a tMSP \citep{bassa14}. This discovery confirmed that the system had shifted to a rotation-powered pulsar state, marking it as one of the few known transitional systems. 

We used ATCA radio data from \citet{hill11}, specifically the values from the reanalysis by \citet{bassa14}, which reported flux densities of 190$\pm$30 $\mu$Jy at 5.5 GHz and 180$\pm$40 $\mu$Jy at 9 GHz, based on observations taken on 2009 November 06. In addition, we stacked all available \textit{Swift} observations of XSS J12270$-$4859 prior to its transition to the pulsar state and used the average spectrum in our SED fitting. We also confirmed that this spectrum is consistent with other X-ray spectra reported in the literature \citep{demartino10,demartino13}.

For the SED fitting, we applied a reddening correction to the optical data using E(B-V) = 0.11 \citep{demartino10}, and we fixed the companion star's temperature within the range [5000--6000] K. Detailed optical studies of XSS J12270$-$4859 in its pulsar state by \citet{riverasandoval18} and \citet{stringer21} both determined the companion star's base temperature to be approximately 5500 K. We also introduced an additional black body component to the model to account for emission from the outer accretion disk. A two-component fit to the infrared-to-ultraviolet data was similarly required in the study of \citet{demartino13}. 

Furthermore, we fixed the indices of the broken power law spectrum for the jet component to $\alpha_{\rm thick}=0.0$ and $\alpha_{\rm thin}=-0.7$, allowing only the break frequency and normalization to be fit. We also fixed the optically thick spectral index of the shock component to $\alpha_{\rm thick}=2.5$, following \citet{baglio23}.

The resulting SED is shown in Fig. \ref{fig:sed_J1227}. Interestingly, XSS J12270$-$4859 exhibits a WISE W3-band detection, which clearly shows excess emission above the levels expected from the companion star and accretion disk. Assuming that the mid-infrared data arise from the compact jet, we constrained the jet break frequency to $\nu_{\rm br} \sim 4-10 \times 10^{13}$ Hz. This value is consistent with those observed in the hard state of XRBs, and, if correct, raises intriguing questions about the large difference in jet break frequencies between XSS J12270$-$4859 and \source.

\subsection{4FGL J1544.5$-$1126}

\begin{figure*}
 \centering
 \includegraphics[width=\linewidth]{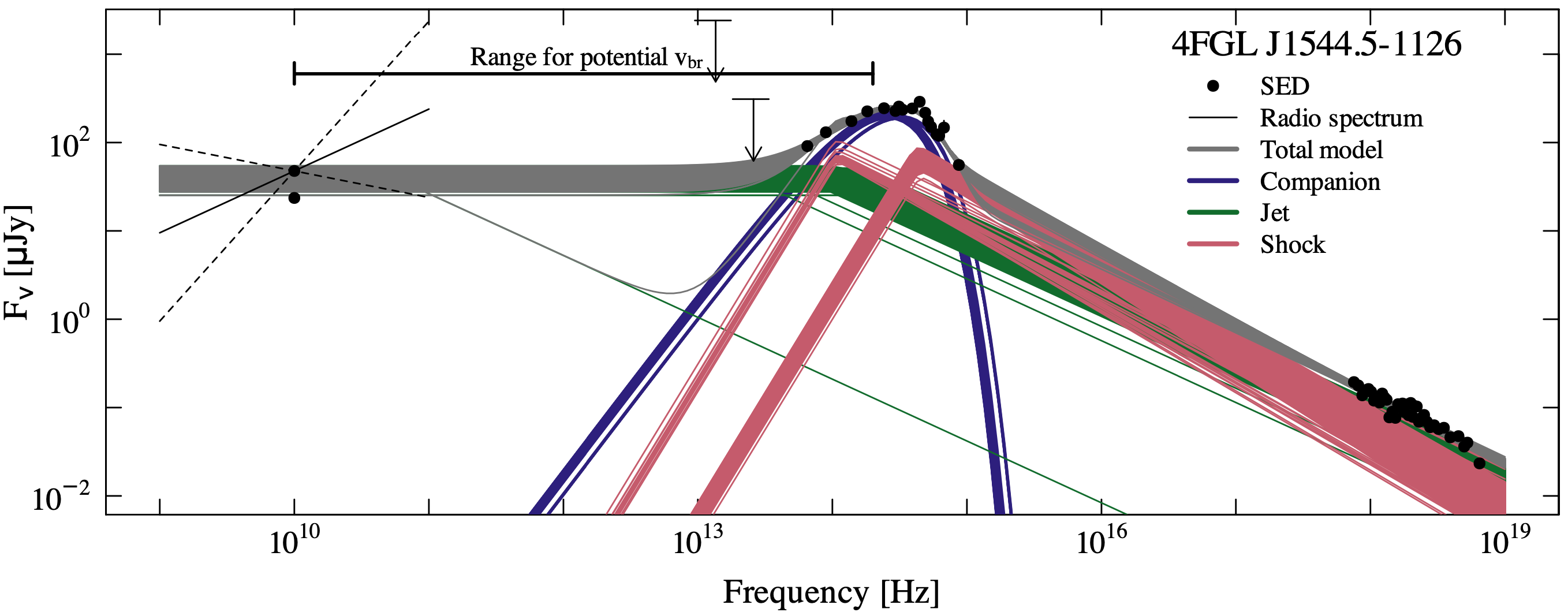}
 \caption{The SED of 4FGL J1544.5$-$1126 is shown as black points. The data is fitted using the same model discussed in Section 3.3 where the ultraviolet data are modeled by the synchrotron shock emission. The downward-pointing arrows represent upper limits from WISE (bands W3 and W4). The solid and dashed black lines indicate the in-band radio spectrum at 10 GHz for the brighter flux observation.}
 \label{fig:sed_J1544}
\end{figure*} 

4FGL J1544.5$-$1126 is a tMSP candidate in a sub-luminous disk state, exhibiting high/low state phenomena in X-rays, similar to confirmed tMSPs. It has been observed across multiple wavelengths, including optical \citep{britt17}, X-rays \citep{bogdanov15}, and radio \citep{jaodand21}. Its companion is likely a K6-7V star with a main sequence mass of 0.7 solar masses and a temperature of 4100$\pm$100 K \citep{britt17}, in a very low-inclination orbit \citep[5--8 deg;][]{britt17}.

We used Very Large Array (VLA) radio data from \citet{jaodand21}. The data were taken on MJD 57154 and MJD 57157, with measured radio flux densities of 23.6 $\pm$ 4.8 $\mu$Jy and 47.7 $\pm$ 6.0 $\mu$Jy at 10 GHz, and an in-band radio spectral index of $\alpha$ = 0.7 $\pm$ 1.0 ($F_{\nu}\propto\nu^{\alpha}$). For the X-rays, we analyzed NuSTAR data as presented in \citet{bogdanov16}, following a similar approach to that described in Section 2.1.2.

In the SED fitting, we corrected the optical data for reddening using E(B-V) = 0.15 \citep{britt17}, and fixed the companion star's temperature within the range [3800--4400] K. We also fixed the spectral indices of the broken power law component corresponding to the jet spectrum to $\alpha_{\rm thick}=0$ and $\alpha_{\rm thin}=-0.7$, and set $\alpha_{\rm thick}=2.5$ for the broken power law component associated with the shock, similar to the model used for XSS J12270$-$4859. 

The resulting SED is shown in Fig. \ref{fig:sed_J1544}. While in most model iterations the jet break frequency is located around $\sim10^{14}$ Hz, we find this result questionable, as the small excess seen in the WISE data could potentially be explained by emission from the companion star.

\subsection{PSR J1023$+$0038}

\begin{figure*}
 \centering
 \includegraphics[width=\linewidth]{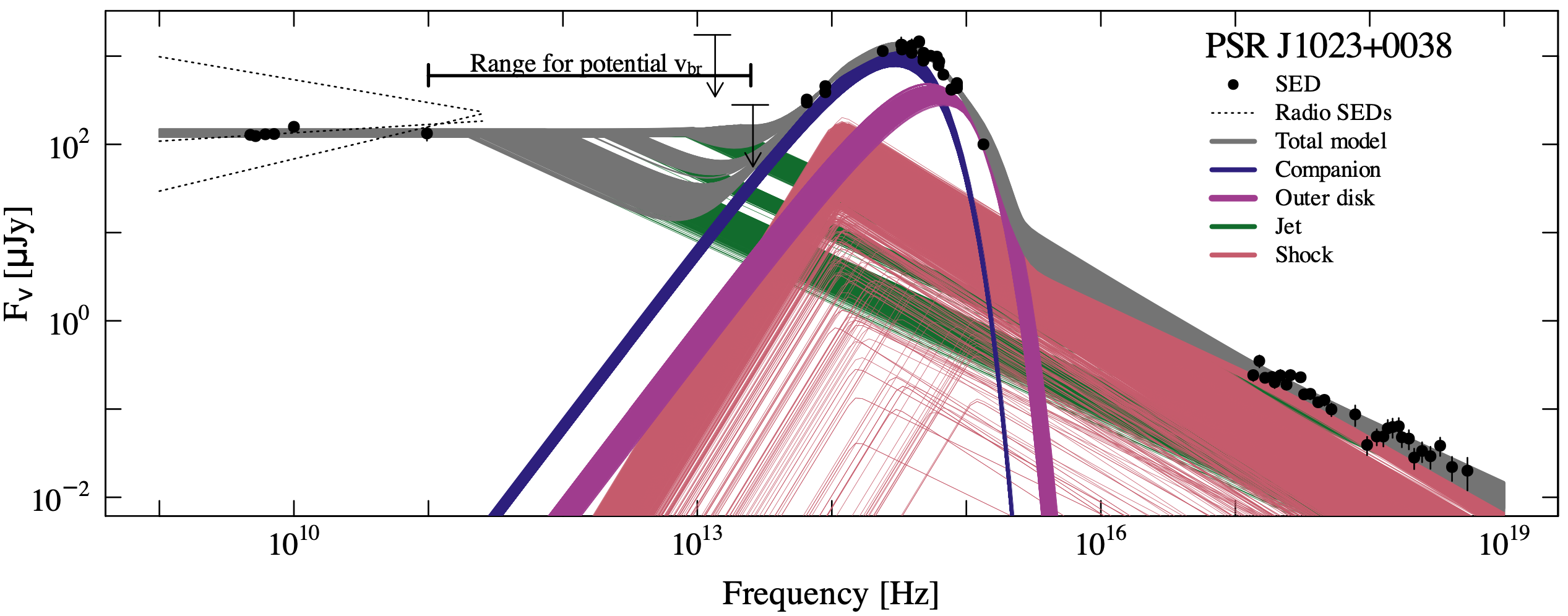}
 \caption{The SED of PSR J1023$+$0038 in the low-mode of the sub-luminous disk state is shown as black points. We fitted the SED using the same model as that for XSS J12270$-$4859. The downward-pointing arrows represent upper limits from WISE (bands W3 and W4). The dashed black lines show the linear fits to the radio data in different modes (high mode has the lowest and flaring mode the highest radio fluxes), as presented in \citet{deller15}. The extent to which the X-ray spectrum can be attributed to the jet spectrum, rather than the shock spectrum, will influence the position of the estimated break frequency (assuming the same optically thin spectrum).}
 \label{fig:sed_J1023}
\end{figure*} 

PSR J1023$+$0038 is the most extensively studied tMSP, with simultaneous multiwavelength data presented by \citet{baglio23}. We use their low-mode dataset, which spans from radio to X-rays, and supplement it with additional radio data from \citet{deller15}, as well as optical and infrared data from the Vizier catalogue. The companion is a G5 star, with a temperature of 6128 K in the disk state \citep{shahbaz22}.

For the SED fitting, we applied a reddening correction to the optical data using E(B-V) = 0.07 \citep{riverasandoval18}, and fixed the companion star's temperature within the range [5500--6500] K. In \citet{baglio23}, a very similar model was used to fit the low-mode SED, with the assumption that the X-ray emission is an extension of the optically thin part of the jet spectrum. However, confirming this would require determining that the jet break frequency is close to $10^{13}$ Hz. The compiled SED (Fig. \ref{fig:sed_J1023}) does not definitively support or refute this with the available data. Therefore, we conservatively estimate the jet break frequency to lie between $10^{11}$ Hz and $10^{13}$ Hz. 



\bsp	
\label{lastpage}
\end{document}